\RequirePackage{fix-cm}
\documentclass[smallextended]{svjour3}       
\smartqed  
\usepackage{graphicx}
\begin{document}

\title{Gamma Rays from colliding winds in massive binaries
}
\subtitle{Status and prospects}


\author{Gustavo E. Romero     
}


\institute{G.E. Romero \at
              Instituto Argentino de Radioastronom\'\i a (IAR) \\
              Casilla de Correos No. 5, 1894 Villa Elisa\\
              Provincia de Buenos Aires, Argentina\\
              Tel.: +54-221-425-4909 \\
              \email{romero@iar-conicet.gov.ar}           
}

\date{Received: date / Accepted: date}

\maketitle

\begin{abstract}
Binary systems formed by early-type stars with strong winds are known to display variable non-thermal radio emission, thermal X-rays, and, at least in one case (Eta Carina), $\gamma$ rays. Some of these systems are quite eccentric and the conditions for efficient particle acceleration and $\gamma$-ray production might manifest only occasionally.    
In this paper I briefly review the physics of colliding wind binaries with emphasis on the observational signatures of non-thermal particle acceleration. I discuss, in particular, the case of the system HD 93129A which is made up by an O2 If* star (the primary) and an O3.5 V star (the secondary). The primary is among the earliest, hottest, most massive, luminous, and windy O stars in the Galaxy. The periastron passage during 2018 will offer an outstanding observational window that will be exploited by an international multi-wavelength campaign. 
\end{abstract}

\section{Introduction}
\label{intro}

Early-type massive stars are very luminous and generate strong winds with mass loss rates up to $\sim 10^{-5}$ $M_{\odot}$ yr$^{-1}$. At the end of their lives the most massive stars can lose mass at even higher rates. The escape velocity of these winds reaches several times $1000$ km s$^{-1}$. 

If these stars form binary systems, something that is quite common since they are born in groups inside giant molecular clouds, the winds will collide. The result of such collision is the formation of a system of shocks, which will heat up the plasma of the winds. When the shocks are adiabatic, diffusive non-thermal particle acceleration can occur (Eichler \& Usov 1993). The resulting population of relativistic particles will cool locally mainly through synchrotron radiation and inverse Compton (IC) up-scattering of stellar photons. As a consequence, colliding wind binaries (CWBs) may become non-thermal radio and even $\gamma$-ray sources (Benaglia \& Romero 2003). The power law index of the spectral distribution of particles can vary widely from $-1$ to values even steeper than $-2$ depending on the system geometry and escape conditions (e.g. Bykov 2014).

De Becker \& Raucq (2013) have identified 43 binary or multiple systems with non-thermal emission in the Galaxy (see also De Becker et al. 2017). Current interferometric radio observations of some of these systems allow to map the evolution of the non-thermal region along the orbit yielding the possibility to get insights on the mechanism of particle acceleration in changing physical conditions.  Thus, CWBs are a kind of natural laboratories for the investigation of cosmic ray production.

\section{Particle acceleration in the colliding winds of early-type binaries}
\label{sec:1}

In an early-type binary  the winds from the primary  and the secondary stars flow nearly radially and
collide at a point located at a distance $r_{i}$ from the $i$-star, given by

\begin{equation}
r_1=\frac{1}{1+\eta^{1/2}}D,\;\;\;\;\;r_2=\frac{\eta^{1/2}}{1+\eta^{1/2}}D.
\end{equation}
Here the subscript ``1'' stands for the primary star, and ``2'' for the secondary, $D$ is the binary separation, and the parameter $\eta$ is defined in terms of the wind terminal
velocities $v_{\infty}$ and the stellar mass loss rates $\dot{M}$:
\begin{equation}
\eta=\frac{\dot{M_2}v_{\infty,2}}{\dot{M_1}v_{\infty,1}}.
\end{equation}

\begin{figure}
  \includegraphics[width=0.75\textwidth]{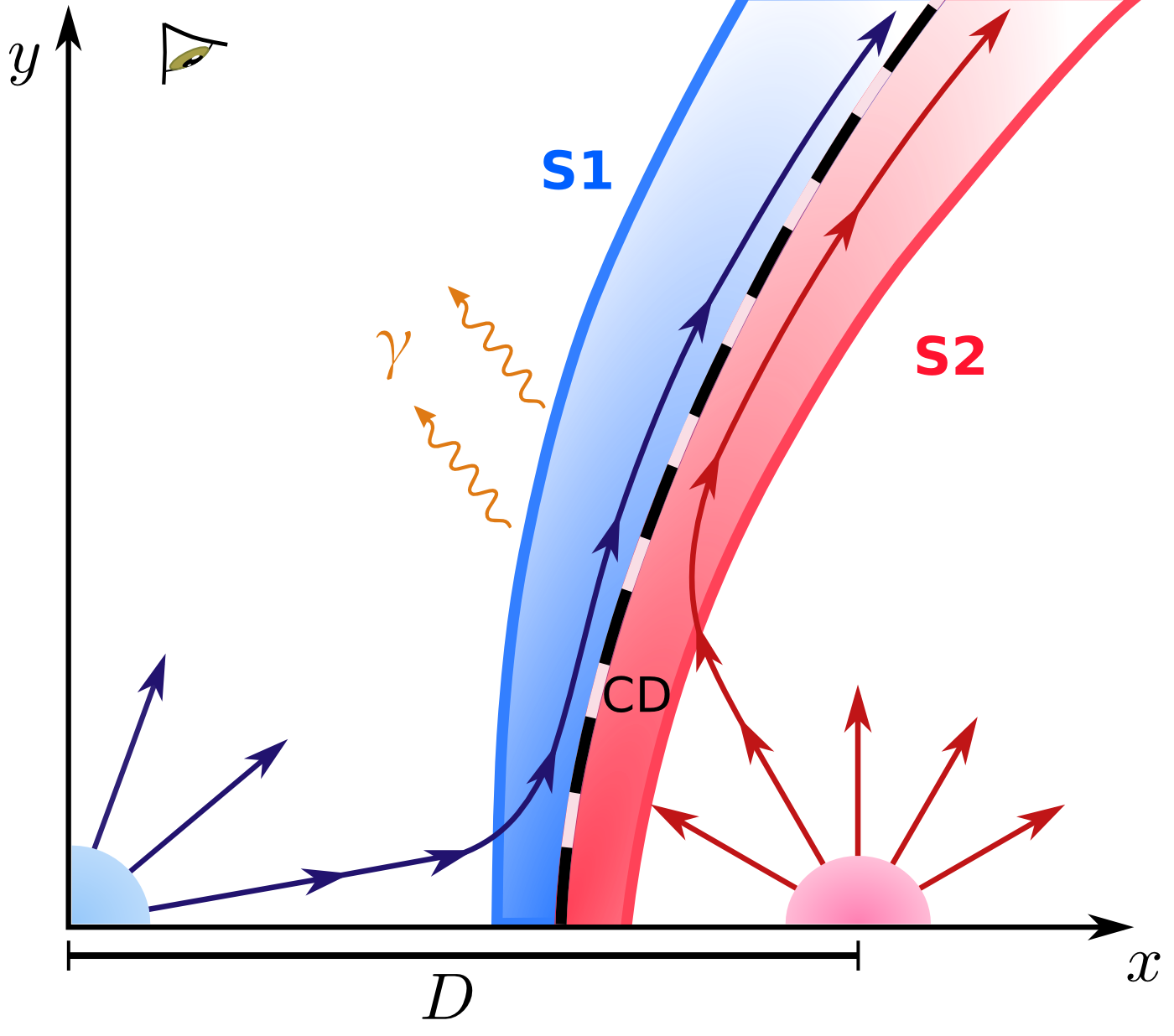}
\caption{Sketch of the system of shocks formed in a CWB. There are two shocks (S1 and S2) that move in opposite directions from the contact discontinuity (CD). From del Palacio et al. (2016).}
\label{fig:1}       
\end{figure}

A system of two shocks is formed in the collision: one moves in the wind of the primary (S1) and the other in the wind of the secondary (S2). Matter flows away along the shocked region that surrounds the contact discontinuity (CD), see Fig. \ref{fig:1}.

Diffusive shock acceleration operates in both shocks if they are adiabatic, so some particles depart from the Maxwell-Boltzmann distribution and acquire a power law spectrum with index $\sim -2$. The cut off of this population will depend on the competing rates of acceleration and losses  (see De Becker 2007 for a detailed discussion).  

In the steady state the particle distribution can be obtained from the transport equation:
\begin{equation}
 \frac{\partial}{\partial E}\biggl[\frac{{\rm d}E}{{\rm d}t}{\bigg\arrowvert}_{\rm loss}N(E)\biggr]+\frac{N(E)}{t_{\rm esc}} = Q(E),
\label{tra}
\end{equation}
where $t_{\rm esc}$ is the convection timescale, $Q(E)$ the source term, and $\frac{{\rm d}E}{{\rm d}t}{\bigg\arrowvert}_{\rm loss}$ includes all kind of losses. In several situations of astrophysical interest magnetic field aplification can occur (e.g. Schure et al. 2012 ). This can significantly modfy the maximum energies that can be achieved and the shape of the spectrum.

\section{Non-thermal radiation}
\label{sec:2}
Relativistic electrons cool in the colliding wind region through synchrotron, Bremsstrahlung, and IC radiation. Protons suffer $pp$ collisions, but they are convected away before cooling. They can reach multi-TeV energies and are later injected into the interstellar medium. As for the electrons, the ratio of magnetic to photon energy densities in the shocked region will determine which mechanism, synchrotron or IC, is more efficient:
\begin{eqnarray}
 \frac{L_\textrm{synchr}}{L_{\textrm{IC}}}=  \frac{U_\textrm{B}}{U_{\textrm{ph}}}.
 \end{eqnarray}
Since the photon density is contributed by the bolometric luminosity of the stars, we get the following expression for the IC power (De Becker 2007):
\begin{eqnarray}
  L_{\textrm{IC}}= \frac{2}{c} \frac{L_\textrm{synchr}}{B^2} \left( \frac{L_1}{r_1^2} + \frac{L_1}{r_1^2} \right),
  \end{eqnarray}
where B is the magnetic field in the colliding wind region and $L_i$ is the bolometric luminosity of the $i$-star.  Conditions vary along the orbit in any eccentric system, so the importance of each contribution will change with the relative position of the stars. The radiation will be affected by different absorption processes that depend on the particle and photon densities, so the appearance of a particular system to distant detectors will change form one case to another (e.g. del Palacio et al. 2016). 

Radiative models of different degrees of complexity have been developed by Benaglia \& Romero (2003), Pittard \& Dougherty (2006), Reimer et al. (2006), and Reitberger et al. (2014), among others. The reader is referred to these works for additional details. 

\section{Current observational status}
\label{sec:3}

Most of particle-accelerating CWBs have been detected in radio using interferometric techniques. A classical example is the well-known system WR140 (Dougherty et al. 2005). In X-rays the detection of non-thermal radiation is hampered by the high temperatures reached by the shocked plasma. This makes the thermal radiation to overwhelm any non-thermal contribution (mainly due to IC) below energies of $\sim10$ keV. So far, only two systems have been shown to have non-thermal X-rays: Eta Carinae and WR 140. 

In $\gamma$-rays the situation is similar: only Eta Carina has been clearly detected by \textit{AGILE} satellite (later confirmed by \textit{Fermi}, see Tavani et al. 2009).  Recently, the association of the Wolf-Rayet binary W11 with a \textit{Fermi} source has been suggested by  Pshirkov (2016) and debated by Benaglia (2016).

\section{Eta Carinae}
\label{sec:4}

Eta Carinae is a heavily obscured and peculiar source. The system includes a  luminous blue variable -- LBV -- star of 90 solar masses and a luminosity of about $5 \times10^6$ $L_{\odot}$. The secondary is a WR star of $\sim$30 solar masses. The mass loss rate of the primary is huge, reaching perhaps up to $10^{-3}$ $M_{\odot}$ yr$^{-1}$. Terminal wind velocities of the primary and secondary winds are estimated to be $\sim 500$ and $3000$ km s$^{-1}$, respectively. For a general overview see Reitberger et al. (2012). 

The conditions in the shocks change along the orbit, inducing variability in particle acceleration, $\gamma$-ray production, and absorption. There is a clear variability with the orbital phase. Different  behaviour is observed at low ($0.3-10$ GeV) and high ($>10$ GeV) $\gamma$-ray energies. Short-term variability occurs at periastron. The flux of the high-energy component varies by a factor $34$ but is different during the two periastron passages observed so far. The multi-wavelength behaviour suggests that the low-energy component is produced by IC scattering whereas the high-energy one might be the result of $pp$-interactions. Maximum energies for protons could reach $\sim10^{15}$ eV (Balbo \& Walter 2017).

\section{HD 93129A}
\label{sec:5}

\begin{figure}
  \includegraphics [width=0.75\textwidth, angle=270]{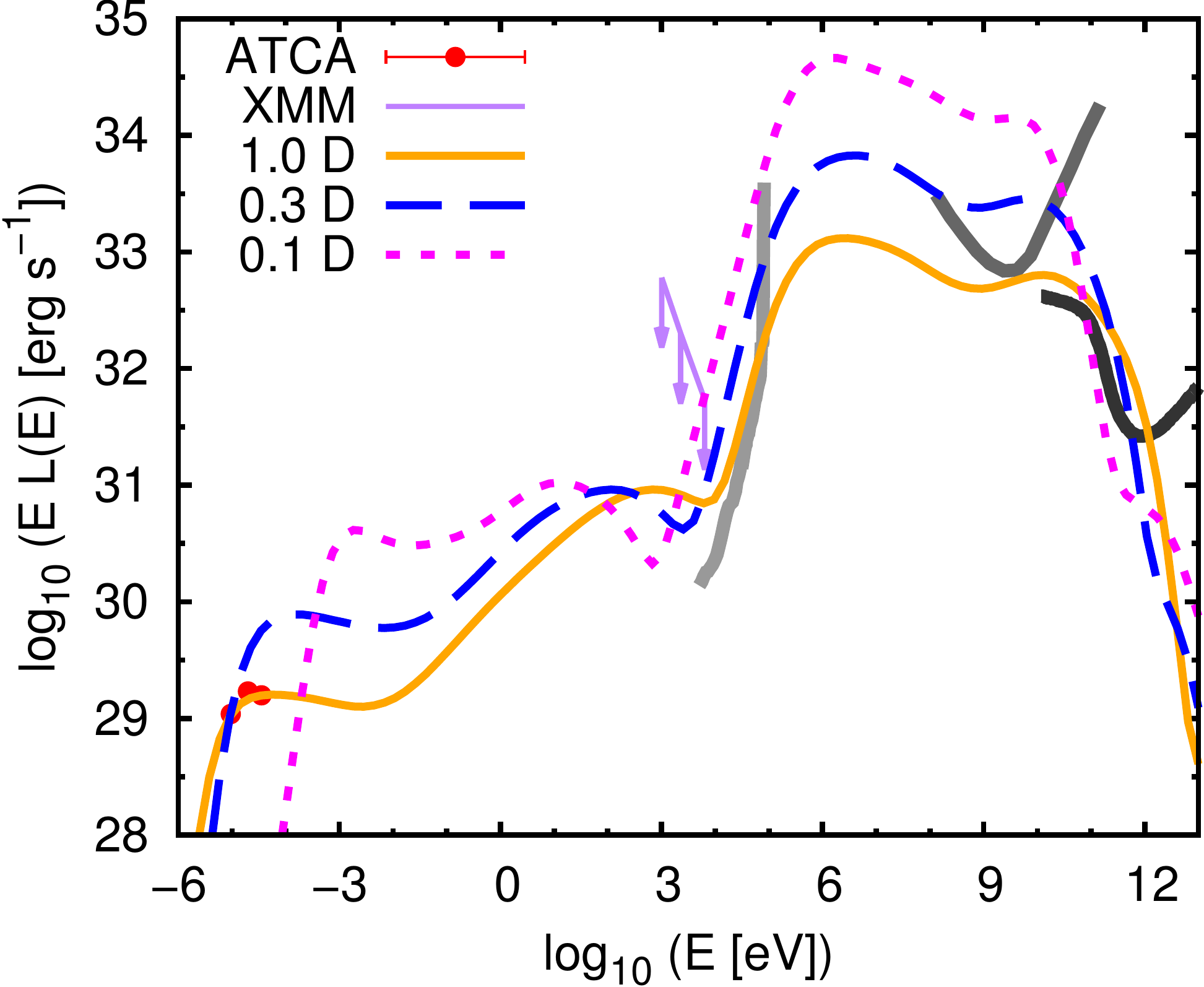}
\caption{Broadband SEDs for the epoch of radio observations (1.0D), and two additionals epochs:
2016 (0.3D), and roughly the periastron passage (0.1D). For the epoch of radio observations, we show the ATCA
data and upper-limits in the X-ray flux derived from
Benaglia et al. (2006).  We show instrument
sensitivity curves for 1-Ms NuSTAR (grey), 4-yr Fermi (dark grey), and
50-h CTA (black). From del Palacio et al. (2016).}
\label{fig:2}       
\end{figure}

The most promising binary system for future $\gamma$-ray detection is perhaps HD 93129A. This is the most massive binary known in the Galaxy, with a total mass of $200\pm45$ $M_{\odot}$, even more than Eta Carinae. The system is located at 2.3 kpc; the primary is an O2 If* star and the secondary is likely an O3.5 V star. The wind collision region of  HD 93129A was recently resolved by the first time using very
large baseline interferometry by Benaglia et al. (2015). These data at radio-frequencies allow to characterize the injection spectrum of relativistic particles, so some detailed models have been produced (del Palacio et al. 2016). A key ingredient is the magnetic field strength. Multifrequency observations including radio, optical, X-ray (\textit{XMM-Newton} and \textit{NuSTAR}), and $\gamma$-ray observatories (\textit{Swift} and \textit{AGILE}) are programmed to cover the periastron passage towards the end of 2018. The observations will allow to investigate the physical properties of the acceleration region and to detect the expected high-energy emission. In Fig. \ref{fig:2} we show the spectral energy distribution (SED) estimated for a low value of the magnetic field and a hard injection spectrum (see del Palacio et al. 2016 for details). The actual SED will allow to solve the inverse problem and to determine constraints on the physical parameters. 

\section{Conclusions}

Colliding wind binaries are, in many cases, non-thermal radio sources. This clearly indicates that particles (at least electrons) are accelerated in them up to relativistic energies. Since this acceleration occurs in a region that has an important photon energy density provided by the luminous stars, IC interactions are unavoidable. However, the high magnetic field can quench the high-energy emission if synchrotron radiation becomes the main cooling channel. In addition, the non-thermal radiation can be suppressed by the high densities of thermal gas and the absorbing photon fields. The conditions for particle acceleration vary along the orbit, with optimal windows for observations occurring at different times for different wavelengths.
 
So far $\gamma$-rays have been detected for sure only from Eta Carinae, an extremely massive colliding wind system that includes a rare luminous blue variable star and a Wolf-Rayet companion. Two different components in the high-energy radiation are distinguished, suggesting that more than a single mechanism is operating in the system. 

HD 93129A is a another very massive colliding wind binary with a long period ($\sim 100$ yr). It is a confirmed non-thermal radio source and an excellent candidate for GeV $\gamma$-ray emission. The forthcoming periastron passage in late 2018 will offer a unique opportunity to study this remarkable system and determine the physical conditions in the acceleration zone. There is an on-going international campaign that includes \textit{Chandra}, \textit{NuSTAR}, \textit{XMM-Newton}, and \textit{Swift}. Data from  \textit{AGILE} would be essential to complement this combined effort\footnote{\textit{Fermi}, being an instrument for surveys, is perhaps not the best option for tracking this event.}.

\begin{acknowledgements}
I thank S. del Palacio, V. Bosch Ramon, P. Benaglia, and M. De Becker for many discussions on this topic. My work was supported by the Helmholtz Association through a
Helmholtz International Fellow Award. Additional
support was provided by the Argentine agency CONICET
(PIP 2014-00338) and the Spanish Ministerio de Econom\'\i a
y Competitividad (MINECO/FEDER, UE) under grant
AYA2016-76012-C3-1-P.
\end{acknowledgements}


\begin{thebibliography}{}
%
%

\bibitem{Ba} 
Balbo, M., \& Walter, R.  (2017) Fermi acceleration along the orbit of Eta Carinae. A\&A 603: A111, 11 pp.
\bibitem{Ben1} 
Benaglia, P., Koribalski, B., \& Albacete Colombo, J. F. (2006) Radio detection of colliding wind binaries. PASA, 23: 50-63.
\bibitem{Ben2}
Benaglia, P.,  Marcote, B.,  Mold\'on, J. et al. (2015) A radio map of the colliding winds in the very massive binary system HD93129A. A\&A 579: A99, 9 pp.
\bibitem{B}
Benaglia, P. (2016) Is the stellar system WR 11 a gamma-ray source? Publications of the Astronomical Society of Australia, 33: id.e017, 9 pp.
\bibitem{BR}
Benaglia, P., \& Romero, G. E. (2003) Gamma-ray emission from Wolf-Rayet binaries. A\&A, 399: 1121-1134.
\bibitem{By}
Bykov, A. M. (2014) Nonthermal particles and photons in starburst regions and superbubbles. A\&ARv, 22: id.77, 54 pp.
\bibitem{DB}
De Becker, M. (2007) Non-thermal emission processes in massive binaries. A\&ARv, 14: 171-216.
\bibitem{DBR} 
De Becker, M. \& Raucq, F. (2013) Catalogue of particle-accelerating colliding-wind binaries. A\&A 558: A28, 20 pp.
\bibitem{DB2}
De Becker, M.,  Benaglia, P.,  Romero, G.E., \&  Peri, C.S. (2017) An investigation into the fraction of particle accelerators among colliding-wind binaries: Towards an extension of the catalogue. A\&A 600: A47, 8 pp.
\bibitem{dP}
del Palacio, S., Bosch-Ramon, V.,  Romero, G.E. \&  Benaglia, P. (2016) A model for the non-thermal emission of the very massive colliding-wind binary HD 93129A.  A\&A 591: A139, 11 pp.
\bibitem{Do}
Dougherty, S. M., Beasley, A. J., Claussen, M. J., Zauderer, B. A., \&  Bolingbroke, N. J. (2005) High-resolution radio observations of the colliding-wind binary WR 140. ApJ 623: 447-459.
\bibitem{EU}
Eichler, D., \& Usov, V. (1993) Particle acceleration and nonthermal radio emission in binaries of early-type stars. ApJ, 402: 271-279.
\bibitem{P}
Pshirkov, M. S. (2016) The Fermi-LAT view of the colliding wind binaries. MNRAS 457: L99-L102.
\bibitem{PD}
Pittard, J. M., \& Dougherty, S. M. (2006) Radio, X-ray, and γ -ray emission models of the colliding-wind binary WR140. MNRAS 372: 801-826
\bibitem{Reimer1}
Reimer, A., Pohl, M., \& Reimer, O. (2006) Nonthermal high-energy emission from colliding winds of massive stars. ApJ 644: 1118-1144.
\bibitem{Reit0}
Reitberger, K., Reimer, O.,  Reimer, A., et al. (2012) Gamma-ray follow-up studies on $\eta$ Carinae. A\&A 544: A98, 9 pp.
\bibitem{Reit}
Reitberger, K., Kissmann, R., Reimer, A., \& Reimer, O. (2014) Simulating three-dimensional nonthermal high-energy photon emission in colliding-wind binaries. ApJ 789: id. 87, 19 pp.
\bibitem{S}
Schure, K. M., Bell, A. R., O'C Drury, L., \& Bykov, A. M. (2012) Diffusive shock acceleration and magnetic field amplification. Space Science Reviews, 173: 491-519.
\bibitem{T}
Tavani, M., Sabatini, S., Pian, E., et al. (2009) Detection of gamma-ray emission from the Eta-Carinae region. ApJ 698, L142-L146.



\end{thebibliography}


\end{document}